\documentstyle[aas2pp4]{article}

\lefthead{Chugai}\righthead{Probing supernova ejecta}

\begin{document}

\title{Probing supernova ejecta by H$\alpha$ damping wings}

\author{Nikolai N. Chugai}
\affil{Institute of Astronomy, RAS, Pyatnitskaya 48, 109017 Moscow,
Russia}

\begin{abstract}
It is predicted that the H$\alpha$ emission line
at the early nebular epoch of type II-P supernovae may display robust
  observational effects of radiation damping wings.
Monte Carlo simulations demonstrate that
 the H$\alpha$ line with the optical depth typical of the early
  nebular epoch acquires extended wings and redshift and becomes
  broader compared to the Sobolev approximation case.
The strength of these effects may be used for constraining 
  parameters of the line-emitting zone.
The anomalous redshift, width and red wing of the H$\alpha$ emission 
  in the supernova SN~1997D on day 150 are explained in terms of
  damping wings effects.
\end{abstract}

\keywords{line: profile, formation --- supernovae: individual (SN 1997D)}

\section{Introduction}

The ejecta structure of a type II-P supernova (SN~II-P) may be 
  sketchy portrayed as a combination of a hydrogen envelope 
  and a "mixed core".
The latter, being a macroscopic mixture of newly 
  synthesised elements, such as, He, O, $^{56}$Ni with 
  a fraction 
 of the hydrogen envelope, is responsible for the luminosity 
 at the nebular epoch, i.e., at the age $>100$ d.
The model of mixed core of SN~II-P is fundamental to the assessement
 of the nuclesynthesis yields, which is the subject of great
 interest. 
Two recent nebular models of SN~1987A 
  (De Kool, Li \& McCray \cite{dk98}; Kozma \& Fransson \cite{kf98})
  demonstrate that the basic physical processes in the mixed core 
 are well understood to make feasible the composition analysis.
However, to achieve this goal a nebular model of SN~II-P should, 
 first, pass the rigorous test on reproducing fluxes and profiles of 
 hydrogen lines, including H$\alpha$.

Here I argue that the model of the H$\alpha$ line in SN~II-P at the 
 early nebular epoch (100--200 d after the outburst) should 
 essentially include effects of radiation damping wings always 
 ignored before.
The physical reason for the damping wings to play a role is
 a high degree of a non-thermal excitation of hydrogen in 
 the mixed core of SN~II-P, which results in the large 
 H$\alpha$ optical depth at the early nebular epoch.
In this situation to model H$\alpha$ damping wing effects the
 Sobolev (local escape) approximation must be abandoned.
The radiation transfer in the physical and the frequency space 
 simulated with the Monte Carlo method is applied to demonstrate 
 expected effects of the H$\alpha$ damping wings in SN~II-P (Section 2).
Recently found in SN~1997D a disparity between the large asymmetry 
  of H$\alpha$ at the early nebular phase and the symmetric H$\alpha$ 
  profile at the late nebular epoch 
 (Chugai \& Utrobin \cite{cu99}) admits a simple explanation in 
 terms of the damping wings (Section 3).
Another motivation to address the issue of the H$\alpha$ damping wings
 in SN~II-P is that this factor may provide an independent diagnostic 
 tool for constraining parameters of the mixed core. 

\section{Expected effects of H$\alpha$ damping wings in SN~II-P}

The Sobolev approximation widely used to 
 analyse supernova optical spectra is 
 justified for most lines because of a highly supersonic expansion,
 $v/u\gg1$, ($v$ and $u$ being the expansion and the thermal velocity, 
 respectively).
However, if the optical depth in damping wings is large, 
 the "supersonic" criterion for the local 
 escape of photons should be replaced by the inequality $v/u\gg a\tau$ 
  (Chugai \cite{nch80}), where $a$ is the Voigt parameter, 
  $\tau=k\lambda t$ is the Sobolev optical depth 
 ($k$ is the line integrated absorption coefficient,
  $\lambda$ is the wavelength, $t$ is the expansion time).
The H$\alpha$  Voigt parameter is
  $a=(A_{32}+A_{31}+A_{21})/(4\pi\Delta \nu_D)
 =3.27\times10^{-3}$,
 where $A_{ik}$ is the Einstein transition probability,
 $\Delta \nu_D$ is the thermal Doppler width (kinetic temperature
 of 5000~K is adopted).
At the early nebular epoch ($\sim 150$ d) the Sobolev optical
 depth of the mixed core in H$\alpha$ may exceed $10^5$
 (Xu et al. \cite{xu92}) implying that $v/u\sim 2\times 10^2$, i.e.,
 is of the order of $a\tau$.
In this case the local escape approximation is too crude, so that
 the full radiation transfer should be applied to compute the
  H$\alpha$ line profile.

One needs, first, to specify the frequency redistribution function for
 the H$\alpha$ damping wings.
Let $x'$ and $x$ be the incident and scattered photon frequencies in the
 atom frame.
With lower and upper levels broadened by the radiation damping, the 
 atomic angle-averaged frequency redistribution function in
 far wings ($|x'|\gg a$) may be written as 
 $r(x',x)=\phi(x')p(x',x)$ (Oxenius \cite{ox86}), where
 $\phi(x)=(a/\pi)/(x^2+a^2)$ is the normalized absorption coefficient
  and 

\begin{equation}
p(x',x)=\frac{a_3}{a}\frac{2a_2/\pi}{(x-x')^2+4a_2^2} +
\frac{a_2}{a}\phi(x).
 \label{eq:prob}
\end{equation}

\noindent Here $a_2=A_{21}/(4\pi\Delta \nu_{\rm D})$ and
 $a_3=(A_{31}+A_{32})/(4\pi\Delta \nu_{\rm D})$ are Voigt parameters
 for the lower and the upper level.
Thermal velocity effects in the wings 
  are negligible and will be ignored here.
This implies that the frequency redistribution
  function in the wings is the same for the atom and the fluid frame.
The redistribution function may be essentially simplified 
 to reduce the number of scattering in the Monte Carlo computations
 preserving good precision.
Since multiple local scattering in the Doppler core 
  ($|x|<x_{\rm c}\approx 3$)
  eventually ends up with the photon re-emitted in the wings,
  the scatterings in the Doppler core may be omitted by adopting
 the function $p(x',x)$ equal to zero in the core 
 $|x|<x_{\rm c}$. After the re-normalization this function 
 is labeled $P(x',x)$.
For instance, the function $\phi(x)$ in the second term 
 of Eq.(\ref{eq:prob}) transforms into $\psi_2(x)=0.5x_{\rm c}/x^2$ 
 for $|x|>x_{\rm c}$ and $\psi_2(x)=0$ otherwise.
Results are not sensitive to the
  choice of the core width and $x_{\rm c}=3$ is assumed.

The H$\alpha$ photon history is simulated as follows.
The photon is launched with the distribution function $\psi_2(x)$.
When propagating it experiences a redshift in the co-moving frame
 until it randomly scatters or escapes.
If scattered, the frequency of the re-emitted photon
  is randomly choosen according to the modified distribution 
 function $P(x',x)$.
The process continues until the photon escapes.

To demonstrate the role of H$\alpha$ damping wings in SN~II-P 
   let us consider a freely expanding envelope
  ($v=r/t$, $v$ being the velocity at the radius $r$ at the
  expansion time $t$) with a homogeneous line-emitting core 
  ($v\leq v_1$) and a scattering halo ($v_1<v<v_2$).
The Sobolev optical depth of the halo decreases outward as 
  $\tau=\tau_2(1-v/v_2)$.
The core, optically thin in the continuum, may emit a continuum 
  radiation from homogeneously distributed sources.
The velocity and the optical depth in the
 core (subscript 1) and in the halo (subscript 2), respectively,
 are given in Table 1.

The homogeneous core with the low optical depth in the model M1 
 produces a parabolic profile (Fig. 1a) expected for the 
  Sobolev approximation.
When the optical depth is high (models M2 and M3)
  the extended wings emerge, while the line core
  becomes broader and gets redshifted.
All the effects are caused by the radiation transfer in the 
 damping wings. 
Note, that in the case of the Sobolev approximation 
  the models M2 and M3 would have the same parabolic 
   profile as that of the model M1.
A conspicuous kink at the blue slope in the
  model M3 is caused by a self-absorption due to non-local
  scattering.
The models M4 and M5 show how the models M1 and M3 are changed
 when a scattering halo and a continuum core are 
 present (Fig. 1b).
The model M5 gives a realistic impression of damping wings effects 
 in H$\alpha$ at the early nebular epoch of SN~II-P.

How would the inhomogeneity of the hydrogen distribution
 in the mixed core change these results?
Generally, the optical depth in the damping wing 
  depends not only on the local concentration of absorbers, but
  also on its filling factor ($f$), and, 
  in lesser degree, both on the size of inhomogeneities, and on
  their topological properties.
In the limit of small scale inhomogeneities, when
  the mean free path length in the far wing exceeds a typical 
  size of the inhomogeneity, the optical depth in the wing 
  may be reduced to a simple form $\Delta\tau=(a\tau_1/x^2)f|\Delta x|$,
  where $\tau_1$ is the Sobolev optical depth and 
  $\Delta x=(dx/dl)\Delta l$ is the frequency shift in comoving 
  frame at the length interval $\Delta l$.
Under this approximation the H$\alpha$ profile from the inhomogeneous
  core is the same as that from the homogeneous core with
  the "effective optical depth" $\tau_{1,\rm eff}=f\tau_1$, 
The analysis of the H$\alpha$ damping wings 
     thus permits us to estimate the H$\alpha$
  effective optical depth in the mixed core, or, equivalently, 
  $fn_2$ ($n_2$ being the population of the second hydrogen level).
It is noteworthy that the derived value of $fn_2$ is
 independent of the adopted distance of a supernova.

\section{Identification of H$\alpha$ damping wings effects in SN~1997D}

The spectra of the type II-P supernova SN~1997D with a low expansion velocity
 have been obtained at the early ($t\approx150$ d) and the
 late ($t\approx300$ d) nebular epoch by Turatto et al. (\cite{t98}).
The nebular spectrum on day 300 was reproduced in the model
 of a spherically symmetric mixed core with the total ejecta mass
 of 6 $M_{\odot}$ and the kinetic energy of $10^{50}$ erg
 (Chugai \& Utrobin \cite{cu99}).
However, it was stressed there that H$\alpha$ on
 day 150 shows a significant redshift in a disparity
 with a spherically symmetric model.
Note, that the asphericity of the $^{56}$Ni distribution might 
 produce the H$\alpha$ asymmetry on day 150 likewise it was in SN~1987A.
However, this reason for SN~1997D is discarded by the absence of H$\alpha$
 asymmetry on day 300.
We will see below that damping wings effects may
  resolve the revealed controversy.

The assumed envelope model is similar to that of the  Section 2, i.e., 
  it consists of a line-emitting homogeneous core and a scattering halo.
For models considered here the core velocity is $v_1=650$ km s$^{-1}$ 
  and the outer velocity of the scattering halo is $v_2=2000$ km s$^{-1}$;
  the latter is consistent with the steep density drop in the range 
  $v>1500$ km s$^{-1}$ for the hydrodynamical model of SN~1997D
  (Chugai \& Utrobin \cite{cu99}).
The Sobolev optical depth in the halo is assumed to decrease outward as 
  $\tau_2(1-v/v_2)$ with $\tau_2=10$.
An additional factor implemented into the model is the Thomson scattering, 
  although it is of minor importance.
The corresponding optical depth ($\tau_{\rm T}$) of the mixed core,
  estimated from the H$\alpha$ luminosity (distance $D=13.43$ Mpc) using
  data by Turatto et al. (\cite{t98}), is 0.16 on day 150 and 0.08
 on day 300. 
Bellow, I adopt $\tau_{\rm T}=0.1$ for $t=150$ d and
 $\tau_{\rm T}=0.05$ for $t=300$ d to roughly take into account that 
 the hydrogen filling factor in the mixed core is essentially less than
 unity (Chugai \& Utrobin \cite{cu99}).

In the late time spectrum on day 300 
 the model without damping wings is not
 very much different to the observed profile (Fig. 2). 
It implies that the damping wings at this epoch are not significant,
 although, the model with the damping wings and
  $\tau_{1,\rm eff}=5\times10^{3}$ fits the observed profile better.
On day 150 the model without damping wings is strikingly dissimilar 
 to the observed profile: the latter is broader and shows both 
 significant redshift and red wing.
The model with damping wings included and 
 $\tau_{1,\rm eff}=5\times10^{4}$ turns out to be
 successful in the description of the observed profile.
Note, that with the core velocity set by the late 
 time profile, the primary fitting parameter is the effective 
 optical depth in the mixed core.

The found effective optical depth value being
 combined with the Sobolev optical depth may be used 
  to derive the hydrogen filling factor in the mixed core.
To estimate the Sobolev optical depth
 I adopt 0.002 $M_{\odot}$ of $^{56}$Ni (Turatto et al. \cite{t98})
 in the mixed core ($v\leq 650$ km s$^{-1}$).
The density $\rho\approx 5\times10^{-13}$ g cm$^{-3}$ is
  adopted on day 150 following the model of SN~1997D 
 (Chugai \& Utrobin \cite{cu99}).
The non-thermal excitation of hydrogen may be calculated
 assuming that all the energy of gamma-rays from the
 radioactive decay of $^{56}$Co is uniformly deposited in the 
 mixed core, which is a sound approximation at the early nebular 
 epoch when the mean free path for gamma-rays is small.
The energy deposited in H-rich matter is spent 
 (apart from Coulomb heating), eventually, on the excitation of the 
 second level of hydrogen.
With the depopulation rate controlled by the two-photon decay 
 and the collisional de-excitation I find
 $n_2\approx 1.3\times 10^4$ cm$^{-3}$ on day 150, which leads to
 the Sobolev optical depth of $\approx 1.7\times10^5$.
The latter being combined with the effective
 optical depth ($\tau_{1,\rm eff}=5\times10^4$) results in the 
  hydrogen filling factor of $f\approx 0.3$.
The latter value is quite realistic, although it is a factor 
  of 1.5 larger than the estimate from the nebular 
  model on day 300 (Chugai \& Utrobin \cite{cu99}).
The difference is not dramatic taking into account possible
 uncertainties of both models.
The factor of ten lower value of the effective optical depth 
 in H$\alpha$ on day 300 compared to that on day 150 is 
  consistent with the decrease of the Sobolev optical depth in 
 H$\alpha$ expected for the mechanism of the non-thermal excitation
 (e.g., Xu et al. \cite{xu92}).

Although describing an inhomogeneity of the excited hydrogen in the 
 mixed core in terms of the filling factor seems sensible,
 a possibility exists that between the mixed core and unmixed 
 hydrogen envelope there is a transition layer of strongly excited 
 hydrogen with the filling factor of unity. 
To estimate qualitatively effect of this component
 I introduced in the model on day 150 an additional spherical  
  layer in the velocity range 650--700 km s$^{-1}$ with filling 
  factor of unity,
  negligibly small net emission, and the Sobolev optical depth being 
 equal to the effective optical depth of the mixed core.
Such a choice of optical depth suggests factor of three
 lower hydrogen excitation in the transition layer than in 
 the mixed core. 
The new model is found to produce the same H$\alpha$ profile as 
 in the old model provided the effective optical depth of the mixed 
  core is of $\tau_{1,\rm eff}=4\times10^{4}$, i.e., 20\% lower 
  compared to the old model.
This implies that the value of the H$\alpha$ effective optical depth
  found from the profile modelling is quite robust to reasonable 
  edditing the geometry of the distribution of the excited 
  hydrogen in the mixed core.

Summing up, the H$\alpha$ damping wings effects provide a natural 
 explanation for the dramatic discrepance between the model based on 
 the Sobolev approximation and the observed profile in SN~1997D 
 on day 150.
This conclusion is strengthen by the absence of a reasonable 
 alternative.

\section{Summary and Discussion}

At the early nebular epoch of SN~II-P the H$\alpha$ emission should 
  exhibit observational signatures of damping wings.
Although, the damping wings were never 
  considered before as relevant for supernova optical spectra, 
  the claimed effects are robust and may be easily identified.
We see damping wings effects in SN~1997D spectrum, where they
  are responsible for the redshift, large width 
 and extended red wing of H$\alpha$ on day 150.
The detection of H$\alpha$ damping wings effects in SN~II-P provides 
  a direct estimate of the H$\alpha$ effective optical depth 
 (i.e. the product of the Sobolev optical depth and the hydrogen
  filling factor) in a distance independent way.
We thus obtain a reliable test for nebular models of SN~II-P.
On the other hand, with a confident nebular model in hand,
  we may able to measure the hydrogen filling factor in the
  mixed core.
The price for this knowledge is the abandoning the Sobolev approximation
  and the need for the full radiation transfer treatment 
  of the H$\alpha$ radiation.

Needless to say that the simple model of the H$\alpha$ profile used
  above for SN~1997D admits some improvements.
For instance, one may take the radial dependence
  of hydrogen excitation and filling factor in the mixed core into
 account.
Another modification, might be 
 the inclusion of the absorption of H$\alpha$ quanta in the 
  Paschen continuum.
This may be done if the radiation transfer and the hydrogen excitation are 
  calculated simultaneously.
More detailed description of the hydrogen inhomogeneities is also welcome,
 although the expected change of the value of the effective optical depth 
 would be within 20\%.
 
To recover the damping wings in other SN~II-P it is necessary 
 to use at least two nebular spectra: at 
  the early epoch, (around day 150) and 
  at the late epoch, but before the dust formation, 
 (e.g., between 300 d and 400 d).
The early nebular epoch is most appropriate to extract the information
  about damping wings, while the late time nebular spectrum provides 
  a template to set the core velocity and to
  secure results from a confusing effects of asymmetric
   $^{56}$Ni distribution.
Notably, it is $^{56}$Ni asymmetry that prevent us from
 using a simple symmetric model to study 
  H$\alpha$ damping wings effects in SN~1987A.

\acknowledgments
I am grateful to Bruno Leibundgut for hospitality at ESO and
 to Massimo Turatto for the generous permission to use the 
 spectra of SN~1997D.
This work was supported by RFFI grant 98-02-16404.



\begin{table*}
  \caption{Parameters of demonstration models}
  \bigskip
  \begin{tabular}{lcccc}
  \tableline
  Model & $v_1$  & $\tau_1\times 10^{-4}$  & $v_2$ & $\tau_2$\\
        & (km s$^{-1}$) &      &       (km s$^{-1}$) &      \\
  \tableline
  M1    &  2000  & 0.001      & 2000  & 0 \\
  M2    &  2000  & 2          & 2000  & 0 \\
  M3    &  2000  & 5          & 2000  & 0 \\
  M4    &  2000  & 0.001      & 4000  & 10  \\
  M5    &  2000  & 5          & 4000  & 10  \\
  \tableline
  \end{tabular}
  \label{tab} 
\end{table*}

\clearpage

\begin{figure}
\plotone{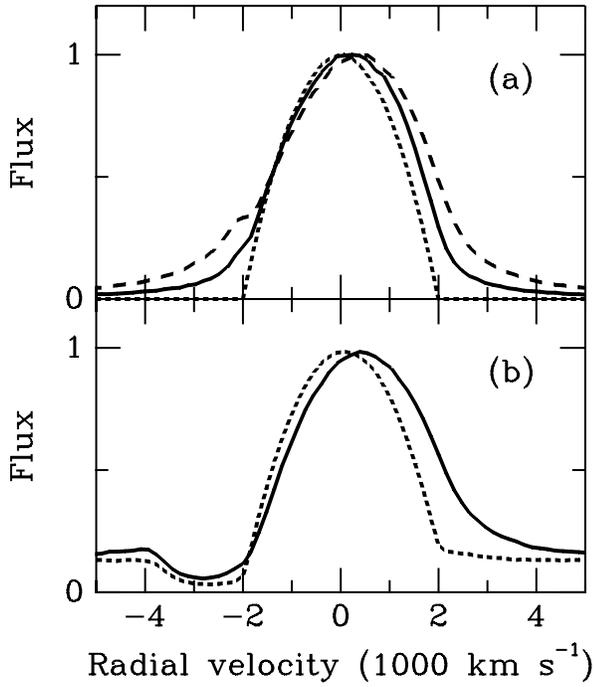}
\caption{Effects of H$\alpha$ damping wings expected for SN~II-P.
The panel ({\em a}) displays the case without continua and a
 scattering halo. The dotted line is the Sobolev approximation  
 (model M1), while the solid line (model M2) and the 
  dashed line (model M3) show effects of damping wings. 
The panel ({\em b}) displays the case with the continua and the
 scattering halo.
The dotted line is the Sobolev approximation (model M4), while 
the solid line shows the model M5 with damping wings.
\label{fig1}}
\end{figure}

\begin{figure}
\plotone{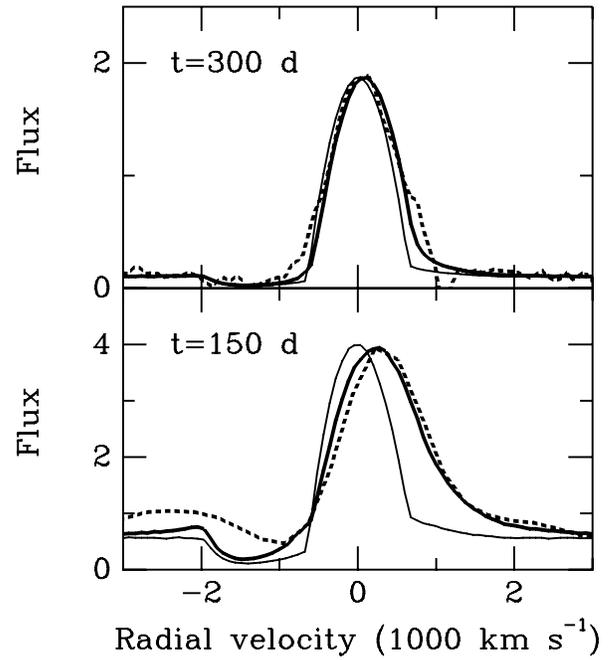}
\caption{Calculated and observed H$\alpha$
in SN~1997D on days 150 and 300. The dotted line is the 
  observed spectrum; the Sobolev approximation is shown by the thin solid
 line, while the thick solid line shows the model with damping wings.
\label{fig2}}
\end{figure}

\end{document}